\documentclass[10pt,conference]{IEEEtran}
\IEEEoverridecommandlockouts
\usepackage{cite}
\usepackage{amsmath,amssymb,amsfonts}
\usepackage{algorithmic}
\usepackage{graphicx}
\usepackage{textcomp}
\usepackage[usenames,tables]{xcolor}
\usepackage{booktabs}
\usepackage{multirow}
\usepackage{xspace}
\usepackage{xparse}
\usepackage{caption}
\usepackage{subcaption}
\usepackage{footnote}
\usepackage{url}
\usepackage[most]{tcolorbox}
\usepackage{array}
\definecolor{ABlue}{HTML}{127bca}
\definecolor{LHScolor}{HTML}{555555}
\usepackage[hidelinks]{hyperref}

\def\BibTeX{{\rm B\kern-.05em{\sc i\kern-.025em b}\kern-.08em
    T\kern-.1667em\lower.7ex\hbox{E}\kern-.125emX}}
\begin{document}

\title{Sampling Projects in GitHub for MSR Studies}

\author{\IEEEauthorblockN{Ozren Dabic, Emad Aghajani, Gabriele Bavota}
\IEEEauthorblockA{\textit{SEART @ Software Institute, USI Universit\`a della Svizzera italiana, Lugano, Switzerland}}
}

\newcommand{\ie}{\emph{i.e.,}\xspace}
\newcommand{\eg}{\emph{e.g.,}\xspace}
\newcommand{\etc}{etc.\xspace}
\newcommand{\etal}{\emph{et~al.}\xspace}
\newcommand{\secref}[1]{Section~\ref{#1}\xspace}
\newcommand{\figref}[1]{Fig.~\ref{#1}\xspace}
\newcommand{\listref}[1]{Listing~\ref{#1}\xspace}
\newcommand{\tabref}[1]{Table~\ref{#1}\xspace}
\newcommand{\ghs}{\textsc{GHS}\xspace}

\newcommand{\reposoverall}{735,669\xspace}
\newcommand{\characteristics}{25\xspace}
\newcommand{\languages}{10\xspace}

\newcommand{\urltt}[1]{\texttt{\url{#1}}}

\newcommand*\circled[1]{\tikz[baseline=(char.base)]{
		\node[shape=circle,fill,inner sep=0pt] (char) {\textcolor{white}{#1}};}}

\newcommand{\droptextshadow}[2]{%
    \tikz[baseline,outer sep=0pt, inner sep=0pt]{
    \node[#1!40!black] at (0,-0.1ex) {#2};
    \node[white] at (0,0) {#2};
}%
}
\newcommand{\DOIbox}[1]{
\tcbsidebyside[
        bicolor,
        sidebyside,
        sidebyside adapt=both,
        sidebyside gap=5pt,
        top=0pt,left=0pt,right=0pt,bottom=0pt,
        boxrule=0pt,rounded corners,
        interior style={top color=LHScolor,bottom color=LHScolor!60!black},
        segmentation style={top color=ABlue,bottom color=ABlue!60!black},
]{%
\droptextshadow{LHScolor}{DOI}
}{%
\droptextshadow{ABlue}{\href{http://dx.doi.org/#1}{#1}}
}%
}

\newboolean{showcomments}
\setboolean{showcomments}{true}
\ifthenelse{\boolean{showcomments}}
  {\newcommand{\nb}[2]{
    \fbox{\bfseries\sffamily\scriptsize#1}
    {\sf\small$\blacktriangleright$\textit{#2}$\blacktriangleleft$}
   }
  }
  {\newcommand{\nb}[2]{}
  }
\newcommand\EMAD[1]{\textcolor{violet}{\nb{EMAD}{#1}}}
\newcommand\GABRIELE[1]{\textcolor{olive}{\nb{GABRIELE}{#1}}}

\maketitle

\begin{abstract}
Almost every Mining Software Repositories (MSR) study requires, as first step, the selection of the subject software repositories. These repositories are usually collected from hosting services like GitHub using specific selection criteria dictated by the study goal. For example, a study related to licensing might be interested in selecting projects explicitly declaring a license. Once the selection criteria have been defined, utilities such as the GitHub APIs can be used to ``query'' the hosting service. However, researchers have to deal with usage limitations imposed by these APIs and a lack of required information. For example, the GitHub search APIs allow 30 requests per minute and, when searching repositories, only provide limited information (\eg the number of commits in a repository is not included). To support researchers in sampling projects from GitHub, we present \ghs (GitHub Search), a dataset containing \characteristics characteristics (\eg number of commits, license, \etc) of \reposoverall repositories written in \languages programming languages. The set of characteristics has been derived by looking for frequently used project selection criteria in MSR studies and the dataset is continuously updated to (i) always provide fresh data about the existing projects, and (ii) increase the number of indexed projects. The \ghs dataset can be queried through a web application we built that allows to set many combinations of selection criteria needed for a study and download the information of matching repositories:\\\urltt{https://seart-ghs.si.usi.ch}.
\end{abstract}



\section{Introduction}\label{sec:intro}

The amount of data available in software repositories is growing faster than ever. At the time of writing, GitHub \cite{GitHub} hosts over 80 Million public repositories\footnote{\urltt{https://api.github.com/search/repositories?q=is:public+fork:true}} accounting for over 1 billion commit activities. Such an unprecedented amount of software data represents the main ingredient of many Mining Software Repositories (MSR) studies. 

One of the fist steps in MSR studies consists in selecting the subject projects, \ie the software repositories to analyze in order to answer the research questions (RQs) of interest. Such a step is crucial to achieve generalizability of the findings and ensure that the selected projects result in useful data points for the goal of the study. For example, a study investigating the types of issues reported in GitHub \cite{Bissyande:issre2013} requires the selection of repositories regularly using the GitHub integrated issue tracker. Instead, a study interested in the pull request (PR) process of OSS projects \cite{Zampetti:saner2019} must ensure that the subject systems actually adopt the PR mechanism (\eg by verifying that at least $n$ PRs have been submitted in a given repository). In addition to RQ-specific selection criteria, several studies adopt specific filters to exclude toy and personal projects. For example, previous works excluded repositories having a low number of stars \cite{wen2020empirical}, commits \cite{pecorelli2020developer}, or issues \cite{Bissyande:issre2013}.

Once the selection criteria have been defined, software repositories satisfying them must be identified. Frequently, the search space is represented by all projects hosted on GitHub that, as previously said, are tens of millions. To query such a collection of repositories, developers can use the official GitHub APIs \cite{GitHubAPI} that, however, come with a number of limitations both in terms of number of requests that can be triggered and information that can be retrieved. For example, the GitHub search API allows for a maximum of 30 requests per minute and each request can return at most 100 results. Only searching for some basic information about the public Java repositories hosted on GitHub would require, at the time of writing, $\sim$160k requests ($\sim$88 hours). If additional information is required for each repository (\eg its number of commits), additional requests must be triggered, making the process even more time expensive. Moreover, setting an appropriate value for the selection criteria (\eg a project must have at least 100 commits) without having an overall view of the available data can be tricky. For instance, researchers cannot easily select the top 10\% repositories in terms of number of commits without firstly collecting this information for the entire population. Finally, given a selection criteria, the GitHub search API provides at most the first 1,000 results (through 10 requests). This means there is no easy way to retrieve all matching results for a selection criteria if it exceed this upper bound. 

To support developers in mining GitHub, several solutions have been proposed. Popular ones are GHTorrent \cite{Gousi13} and GHArchive \cite{GHArchive}. Both projects continuously monitor public events on GitHub and archive them. While the value of these tools is undisputed, as the benefits they brought to the research community, they do not provide a handy solution to support the sampling of projects on GitHub accordingly to the desired selection criteria. For example, computing the number of commits, issues, \etc for a repository in GHTorrent would require MySQL queries aimed at joining multiple tables. 

We present \ghs (GitHub Search) \cite{GitHubSearch}, a dataset and a tool to simplify the sampling of projects to use in MSR studies. \ghs continuously mines a set of \characteristics characteristics of GitHub repositories that have been often used as selection criteria in MSR studies and that, accordingly to our experience in the field, can be useful for sampling projects (\eg adopted license, number of commits, contributors, issues, and pull requests). 

The tool behind \ghs can be configured to mine projects written in specific programming languages. As of today, it mined information about over 700k repositories written in \languages different languages (\ie Python, Java, C++, C, C\#, Objective-C, Javascript, Typescript, Swift, and Kotlin). 

A stable version of the dataset is hosted on zenodo \cite{ghs-dataset} and it features \reposoverall repositories written in the previously mentioned languages. As detailed in the following, \ghs has been designed with scalability in mind and to specifically support the sampling of projects for MSR study. While the user can download the dataset and query it with an ad-hoc script, a querying interface with export features is available at {\bf \url{https://seart-ghs.si.usi.ch}}.

\section{The Dataset}

This section describes \ghs \cite{ghs-dataset}, a dataset containing information about \reposoverall GitHub public repositories that can be used by researchers to easily select projects for an empirical study. In particular, \characteristics characteristics of each project are mined, stored, and continuously updated. Our mining tool exploits the GitHub search API and an ad-hoc crawler we built to collect specific information from the repositories' homepage. \tabref{tab:characteristics} lists the collected characteristics, together with a short description for each of them, the source from which the information is mined and one example of works in the literature that used such a characteristic in the empirical study (or ``-'' if we did not find a related reference).

\begin{figure}
	\centering
	\includegraphics[width=0.6\linewidth]{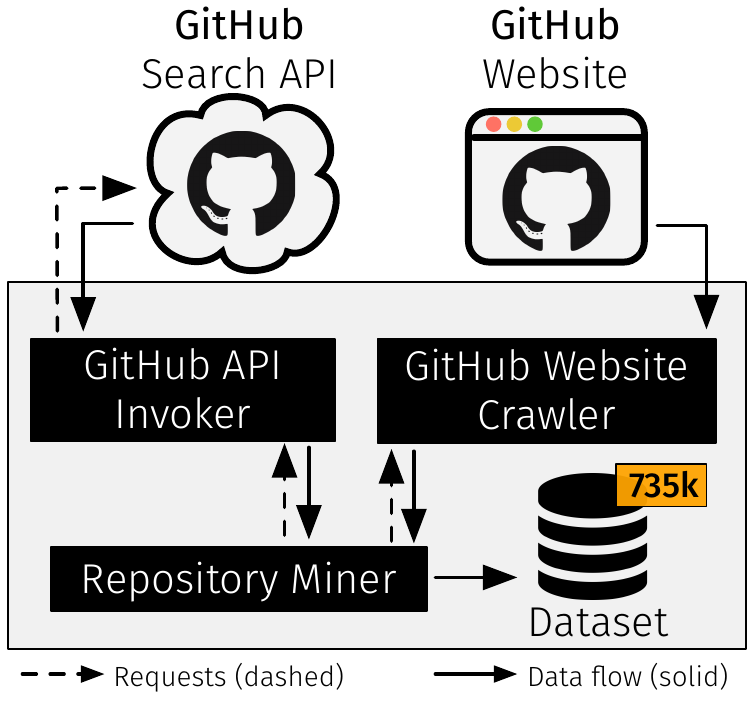}
	\caption{The \ghs architecture}
	\label{fig:process}
\end{figure}

\figref{fig:process} depicts the main steps behind the data collection process put into place to build \ghs. The following subsections detail such a process.

\begin{table*}[h]
\centering
\scriptsize
\resizebox{1\textwidth}{!}{
\begin{tabular}{lllr} 
\toprule
\textbf{Charcteristic} & \textbf{Description} & \textbf{Mining Source}& \textbf{Used in}\\
\midrule
	\texttt{name} & Name of the repository in the form \texttt{user\_name/repo\_name} & GitHub Search API & \cite{muse2020prevalence}\\
	\texttt{commits} & Number of commits on the default branch & Repository's landing page & \cite{gonzalez2020did}\\
	\texttt{last\_commits\_sha} & The SHA-1 hash of the latest commit on the default branch& Repository's landing page & -\\
	\texttt{last\_commits} & The date of the latest commit on the default branch & Repository's landing page & \cite{gonzalez2020state}\\
	\texttt{license} & The license used for the repository (if any) & GitHub Search API & \cite{vendome2017license}\\
	\texttt{branches} & Number of remote branches & Repository's landing page & \cite{Han:compsac2019}\\
	\texttt{default\_branch} & Name of the default branch & GitHub Search API & -\\ 
	\texttt{contributors} & Number of contributors & Repository's landing page & \cite{pecorelli2020developer}\\
	\texttt{releases} & Number of releases \cite{releases} & Repository's landing page & \cite{Moreno:tse2017}\\
	\texttt{watchers} & Number of users watching the repositories & Repository's landing page & \cite{sheoran2014understanding}\\
	\texttt{stars} & Number of stars the repository received & GitHub Search API & \cite{Zampetti:saner2019}\\
	\texttt{forks} & Number of repositories forked from this repository & GitHub Search API & \cite{gonzalez2020state}\\
	\texttt{is\_fork\_project} & Whether the projects is a fork & GitHub Search API & \cite{bryksin2020using}\\
	\texttt{size} & The size of project (in kilobytes) & GitHub Search API & \cite{borrelli2020detecting}\\
	\texttt{created\_at} & Date when the repository is created & GitHub Search API & \cite{bryksin2020using} \\ 
	\texttt{pushed\_at} & Latest date when a commit is pushed to any of the repository's branches& GitHub Search API & \cite{gonzalez2020did}\\
	\texttt{updated\_at} & Latest date when the repository object is updated, \eg description changed& GitHub Search API & -\\
	\texttt{homepage} & The repository's homepage URL (if any)& GitHub Search API & \cite{Aghajani:icse2019}\\
	\texttt{main\_language} & The main language that the repository's source code is written in & GitHub Search API & \cite{nakamaru2020empirical}\\
	\texttt{total\_issues} & Total number of issues (both open and closed issues) & Repository's issues page & \cite{Bissyande:issre2013}\\
	\texttt{open\_issues} & Number of open issues & Repository's issues page & \cite{Bissyande:issre2013}\\
	\texttt{total\_pull\_requests} & Total number of pull requests (both open and closed issues) & Repository's pull requests page & \cite{Zampetti:saner2019}\\
	\texttt{open\_pull\_requests} & Number of open pull requests  & Repository's pull requests page  & \cite{Zampetti:saner2019}\\
	\texttt{has\_wiki} & Whether the repository has wiki & GitHub Search API & \cite{Tantisuwankul:jss2019}\\
	\texttt{archived} & Whether the repository is marked as archived (\ie read-only) & GitHub Search API & \cite{Coelho:esem18}\\
\bottomrule
\end{tabular}
}
\caption{Characteristics stored in \ghs for each GitHub project}
\vspace{-0.5cm}
\label{tab:characteristics}
\end{table*}


\subsection{Data Extraction}

As depicted in \figref{fig:process}, the \ghs data collection process is carried out through three main components.

\textbf{1. GitHub API Invoker}: This component has two main responsibilities. First, it can retrieve the list of repositories (i) written in a specific language, and (ii) created or updated during a certain time period. The latter feature is needed, as detailed later, to overcome the GitHub API maximum result limit of 1,000 results per request. To retrieve, for example,  the list of repositories written in Java and updated in March 2020, the following GitHub API request is triggered: 

\begin{center}
\small
\urltt{https://api.github.com/search/repositories?q=fork:true+is:public+language:Java+created:2020-03-01..2020-04-01}
\end{center}

For each collected repository, the information in \tabref{tab:characteristics} having ``GitHub Search API'' as mining source is retrieved. 

Second, this component is in charge of monitoring if the GitHub access token being used for mining has not exceeded its request limit. Indeed, we use authenticated requests to increase the usage limits imposed by the GitHub API.

\textbf{2. GitHub Website Crawler}:  This component is used to collect, for a given repository, all information in \tabref{tab:characteristics} having a repository's webpage as mining source. Since the information of interest is scattered in different pages, this component mines the repository's (i) landing page \cite{landingPage}, (ii) issues page \cite{issuesPage}, and (iii) pull requests page \cite{prPage}. 

We parse the HTML of these pages by using the CSS selectors containing the information of interest. For this task we primarily rely on the \emph{jsoup} library \cite{jsoup}. Unfortunately, due to the use of dynamic content generation in the GitHub pages, not all elements are present when downloading the content of a page, \eg the number of contributors is dynamically generated, and cannot always be captured using \emph{jsoup} (it depends on the time required for loading the needed information). When \emph{jsoup} fails in retrieving a specific information, we rely on the \emph{Selenium WebDriver for Chrome} \cite{selenium}, which provides the possibility to wait for the required information to load. Since \emph{Selenium} introduces a significant performance drawback, it is only used as backup strategy when \emph{jsoup} returns an error. 

We are aware that mining CSS selectors as a strategy to collect information can require future updates if the GitHub UI substantially changes. We considered such a scenario in our implementation by using, when possible, generic selectors that are unlikely to change over time. Also, this ``maintenance cost'' is counterbalanced by the high performance in retrieving the required information ensured by the webpages parsing. 

\textbf{3. Repository Miner}: This is the core component orchestrating the collection of the \ghs dataset. Before describing how it works, it is important to clarify that the set of programming languages of interest (\ie the ones for which repositories will be mined) is defined by the \ghs administrator. In our case, we set the \languages languages composing the current version of the dataset. The \emph{Repository Miner} implements a mining algorithm that is triggered every six hours for continuously updating the information in \ghs. For each programming language of interest, the algorithm checks if any prior mining has been conducted. If no record of prior mining is found, the \emph{GitHub API Invoker} is triggered to mine all repositories created or updated between January 1\textsuperscript{st} 2008 (GitHub started in February 2008) and the current time minus two hours\footnote{We ignore the last two hours since it takes time for the GitHub's internal database to sync newly created projects.}. If, instead, a previous mining process $MP$ has been performed for the specific language, the \emph{GitHub API Invoker} collects all repositories created or updated between the last date mined in $MP$ and the current time minus two hours.

In both cases the \emph{GitHub API Invoker} collects all repositories (i) written in the selected language, (ii) created/updated during the selected interval, and (iii) having at least 10 stars. The decision of only collecting repositories having at least 10 stars aims at drastically reducing the number of repositories we store and makes the data collection more scalable (\eg from preliminary analyses we performed on Java, $<$5\% of repositories have at least 10 stars). We acknowledge that, as also shown in previous work \cite{Munaiah2017}, the number of stars is not a good proxy for repositories quality or relevance, and there are better ways to automatically identify engineered GitHub projects (\eg the Reaper tool \cite{Munaiah2017}). However, we believe that the 10 stars threshold provides a reasonable compromise between the quality of data and the time required to mine and continuously update all projects. 

If the \emph{GitHub API Invoker} retrieves more than 1,000 repositories for a time interval, it splits the interval in half, and the two new time intervals are pushed to a priority queue handling the requests to process. Such a mechanism is needed since the GitHub API only provides the first 1,000 results for a request. The algorithm recursively picks and process the oldest interval from the queue until it is empty, meaning the mining for the current language is completed. 

Otherwise, if there are less than 1,000 results for an interval, the algorithm iterates over the result list. For each retrieved repository, the algorithm scrapes the missing information from the repository web pages (using the \emph{GitHub Website Crawler}), and saves the full record to a database. Our algorithm can mine/update $\sim$20k repositories everyday.

\subsection{Data Storage}
The data collected for all repositories (\tabref{tab:characteristics}) is stored in a MySQL database. When updated information about a previously mined repository is collected, the corresponding rows for that repository will be updated with the new information (\ie no new row is created). 

While this ensures that the repository data contained within \ghs is kept updated, \ghs does not offer an overview of the historic evolution of said characteristics. 

A stable version of the dataset, exported on January 28\textsuperscript{th} 2021, is hosted on zenodo \cite{ghs-dataset} and it features \reposoverall repositories written in \languages languages. \DOIbox{10.5281/zenodo.4476391}

\begin{figure*}
	\centering
	\includegraphics[width=0.9\linewidth]{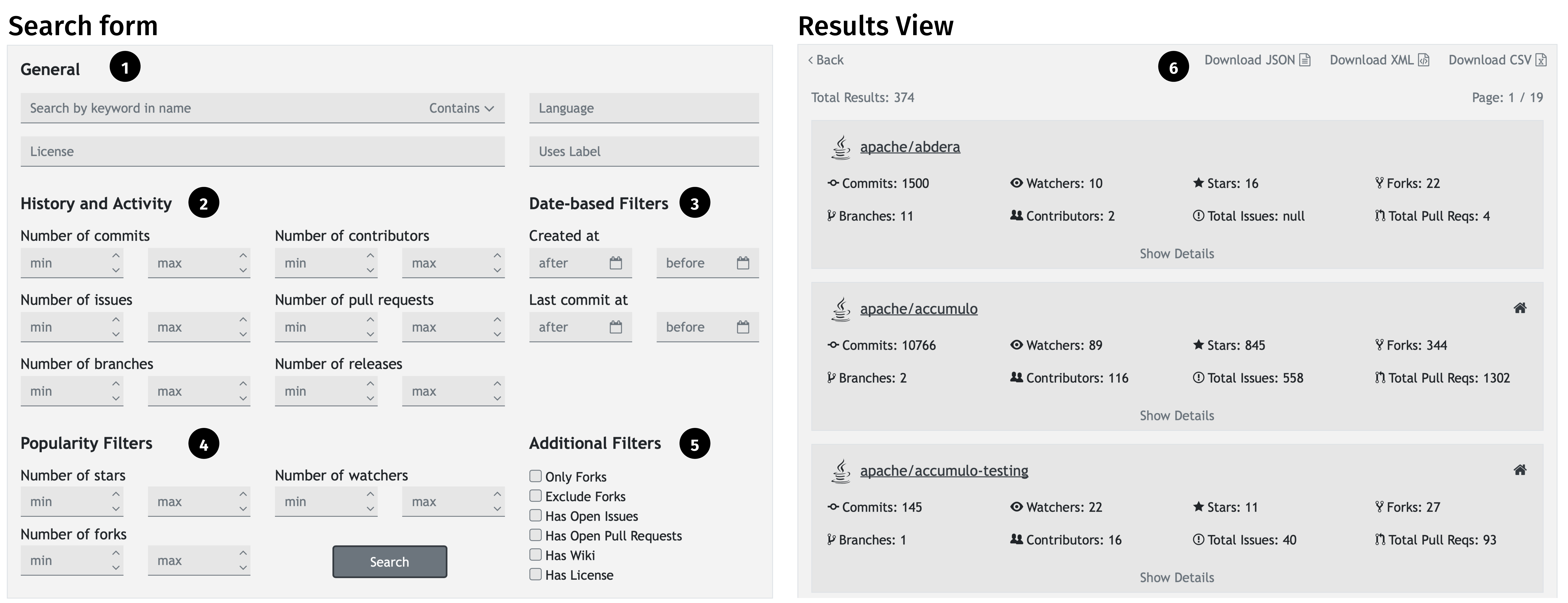}
	\caption{GUI to query \ghs \cite{GitHubSearch} (left) and results page with export options (right).}
	\label{fig:gui}
	\vspace{-0.15cm}
\end{figure*}

\subsection{Querying \ghs}
The latest and continuously growing version of our dataset can be downloaded/queried through our online platform \cite{GitHubSearch}. \figref{fig:gui} depicts the GUI we provide to query \ghs (left part) and an example of results page obtained by searching for the Apache Java repositories having at least 100 commits (right). 

General filters \circled{1} can be applied to select projects containing a specific string in their name (\eg ``apache/'' will return all projects run by the Apache Software Foundation), having a specific license, written in a given language or using specific labels for their issues (\eg ``refactoring''). The latter feature is still under development, which is why we do not present issue labels in the stable version of \ghs. 

Projects can also be filtered based on their history and activity (\eg number of commits, releases) \circled{2}, even only retrieving repositories that had activities in a specific time frame \circled{3}. Finally, filters labeled with \circled{4} concern popularity indicators, while those with \circled{5} allow to further refine the results list by removing, for example, forks.

By clicking on the ``Search'' button, the repositories satisfying the search criteria are shown, giving the possibility to the user to inspect the results list and, eventually, download it in different formats \circled{6}.




\section{Related Work}

To support researchers in MSR, several solutions have been proposed. GHArchive \cite{GHArchive} records the public GitHub activities on an hourly basis as \texttt{json} archives. This is done by mining the GitHub public event stream (\eg a user creating a repository, a repository gaining a new watcher) through the use of webhooks \cite{GitHubAPIWebhook}. This means that, for example, to sample all Java repositories created in 2012 we must retrieve all repositories linked to a ``create'' event from each hour, of each day, of each month of the year. This translates in scanning $\sim$8 thousand files for said events. Thus, while GHArchive is a fantastic data source for MSR studies, it is not convenient for sampling repositories.

GHTorrent \cite{Gousi13} continuously collects data from the GitHub API storing it in both relational and non-relational databases. It likely offers the most used dataset in MSR studies, thanks to the huge amount of stored data and no limitations posed on its querying. However, as mentioned in \secref{sec:intro}, retrieving specific information such as the number of commits in a repository may require formulating queries on quite a large dataset. \ghs, as compared to GHTorrent,  (i) stores only basic repository information needed for making projects' sampling convenient, and (ii) provides a handy GUI to query the dataset.


Software Heritage \cite{dicosmo:hal2017} aims at preserving software in source code form including, \eg projects deleted from GitHub. It contains, at the date of writing, over 150M repositories featuring almost 10B source files. The focus of such a dataset is different from \ghs since Software Heritage is not explicitly meant to simplify projects sampling for empirical studies based on (pre-computed) selection criteria.

Surana \etal \cite{surana2020tool} proposed \emph{GitRepository}, a tool to extract structured information from GitHub repositories related to contributors, issues, pull requests, releases, and subscribers. The authors do not provide a dataset, but a tool able to create a dataset using the GitHub API. \ghs provides a wider variety of information, allowing for a better sampling made easy trough its GUI. In addition to the discussed works, some older projects are no longer active. 

Markovtsev and Long introduced \emph{Public Git Archive} \cite{markovtsev2018public}, a dataset of $\sim$180k repositories having at least 50 stars. The dataset has been released in 2018 and, to the best of our knowledge, is not kept updated. 

Bissyand\'e \etal \cite{Bissyande:2013} presented \emph{Orion}, a corpus of software projects collected from GitHub, Google Code \cite{googleCode} and Freecode \cite{freeCode}. To query \emph{Orion} a custom designed DSL language must be used. The project webpage \cite{orion} is no longer accessible.


\section{Future Work}
There are four main directions in which we are improving \ghs. First, we will add more and more programming languages over time. Doing this is as easy as changing a configuration file. Second, we will finalize the collection of the issue labels that can be used, for example, when a researcher is interested in repositories explicitly using specific labels such as \emph{refactoring} or \emph{documentation}. The GUI already supports such a feature, while the crawling of this information is not yet finalized. Third, the code behind \ghs is open source \cite{githubRepo} and we plan to collect requests for additional project characteristics to include in \ghs from the research community through its issue tracker. Lastly, we will focus on improving performance, especially in terms of data mining.


\section{Conclusions}
We presented \ghs (GitHub Search), a dataset to simplify the sampling of projects for MSR studies. A stable version of \ghs is available on zenodo \cite{ghs-dataset} and features information about \reposoverall GitHub repositories written in \languages languages. The dataset is continuously updated and expanded, with its latest version available at {\bf \url{https://seart-ghs.si.usi.ch}} together with a handy querying interface.

\section{Acknowledgments}
This project has received funding from the European Research Council (ERC) under the European Union's Horizon 2020 research and innovation programme (grant agreement No. 851720).

\bibliography{bibliography}
\bibliographystyle{IEEEtran}

\end{document}